\documentclass[twocolumn]{emulateapj}

\newcommand{\Lsun}{$L_{\odot}$}
\newcommand{\Msun}{$M_{\odot}$}
\newcommand{\Rsun}{$R_{\odot}$}

\shorttitle{V819 Tau: A Rare WTTS with a Weak Infrared Excess}
\shortauthors{Furlan et al.}
\submitted{To appear in ApJ}

\begin{document}

\title{V819 Tau: A Rare Weak-Lined T Tauri Star with a Weak Infrared Excess}


\author{E. Furlan\altaffilmark{1,2}, W. J. Forrest\altaffilmark{3},
B. A. Sargent\altaffilmark{4}, P. Manoj\altaffilmark{3}, 
K. H. Kim\altaffilmark{3}, Dan M. Watson\altaffilmark{3}}

\altaffiltext{1}{Jet Propulsion Laboratory, California Institute of Technology,
Mail Stop 264-767, 4800 Oak Grove Drive, Pasadena, CA 91109; Elise.Furlan@jpl.nasa.gov}
\altaffiltext{2}{{\it Spitzer} Fellow}
\altaffiltext{3}{Department of Physics and Astronomy, University of Rochester, Rochester, 
NY 14627; forrest@pas.rochester.edu, manoj@pas.rochester.edu, khkim@pas.rochester.edu, 
dmw@pas.rochester.edu}
\altaffiltext{4}{Space Telescope Science Institute, 3700 San Martin Drive, Baltimore,
MD 21218; sargent@stsci.edu}

\begin{abstract}
We use {\it Spitzer} data to infer that the small infrared excess of V819 Tau, 
a weak-lined T Tauri star in Taurus, is real and not attributable to a ``companion'' 
10\arcsec\ to the south. We do not confirm the mid-infrared excess in HBC 427 
and V410 X-ray 3, which are also non-accreting T Tauri stars in the same region;
instead, for the former object, the excess arises from a red companion 9\arcsec\
to the east.
A single-temperature blackbody fit to the continuum excess of V819 Tau implies 
a dust temperature of 143 K; however, a better fit is achieved when the weak 
10 and 20 $\mu$m silicate emission features are also included.
We infer a disk of sub-$\mu$m silicate grains between about 1 AU and several
100 AU with a constant surface density distribution. The mid-infrared excess of
V819 Tau can be successfully modeled with dust composed mostly of small 
amorphous olivine grains at a temperature of 85 K, and most of the excess 
emission is optically thin. 
The disk could still be primordial, but gas-poor and therefore short-lived, or already 
at the debris disk stage, which would make it one of the youngest debris disk 
systems known.
\end{abstract}


\section{Introduction}

The evolution and dissipation of protoplanetary disks has been an active area of research
for the last few decades; especially the advent of sensitive near- and mid-infrared
observations has allowed us to explore inner disk regions (out to a few AU), encompassing 
the radii where planets are thought to form. The role of planet formation in disk dissipation
has been all but proven; while disks likely evolve from a flared, optically thick configuration to 
a flat, settled disk via grain growth and settling, it is not clear to what extent the eventual 
dissipation of the remaining dust and gas and the formation of planets are linked. 

A key evolutionary stage in disk dissipation is the transition from the classical T Tauri
stage, when a pre-main-sequence star is surrounded by an accreting disk, to the
weak-lined T Tauri phase, when accretion ends and the disk disappears. This transitional
stage is believed to be short ($\lesssim$ 10$^5$ years), since only few stars with
vanishing infrared excesses have been observed \citep[e.g.,][]{skrutskie90,simon95}.
The Infrared Spectrograph\footnote{The IRS was a collaborative venture between Cornell 
University and Ball Aerospace Corporation funded by NASA through the Jet Propulsion Laboratory 
and the Ames Research Center.} \citep[IRS;][]{houck04} on board the {\it Spitzer Space 
Telescope} \citep{werner04} has provided new insights into the study of transitional disks.
For example, in the nearby and well-studied Taurus star-forming region \citep{kenyon95}, 
out of a total of 111 T Tauri stars, five objects were identified whose inner disk regions are 
cleared to different degrees, but outer, optically thick disks remain, delimited by a well-defined 
inner disk rim \citep{dalessio05,calvet05,furlan06,espaillat07}. Planet formation, 
photoevaporation, or inner disk draining induced by the magneto-rotational instability
may play a role in removing the inner disk \citep{clarke01,quillen04,alexander07,chiang07}, 
but also close binary companions can clear inner regions due to orbital resonances 
\citep{artymowicz94,ireland08}.

\begin{figure}
\centering
\includegraphics[angle=90, scale=0.35]{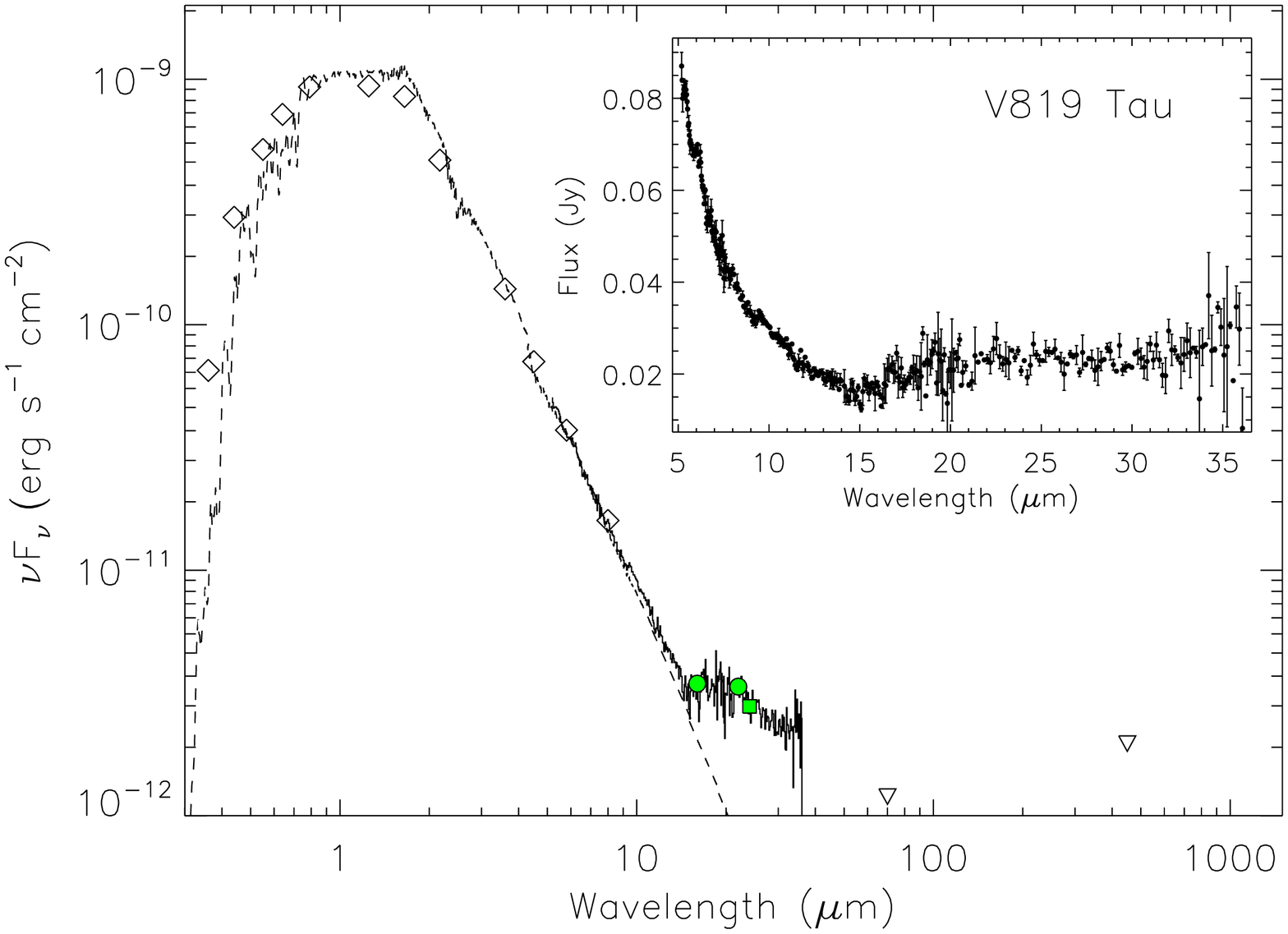}
\caption{The spectral energy distribution of V819 Tau; the optical photometry
is from \citet{kenyon95}, JHK$_s$ photometry from 2MASS \citep{skrutskie06},
IRAC fluxes from \citet{luhman06}, and the 450 $\mu$m upper limit from
\citet{andrews05}. The IRS spectrum (also shown with error bars in the figure
inset), IRS peak-up fluxes, and MIPS fluxes are from this work. The data were 
dereddened using Mathis's reddening law for $R_V$=3.1 \citep{mathis90} and 
$A_V$=1.7 \citep{furlan06}. The stellar photosphere is a scaled NextGen model
with $T_{eff}$=4000 K and $log(g)$=3.5 (see text for details).
\label{V819Tau_IRS}}
\end{figure}

V819 Tau is a weak-lined T Tauri star with a spectral type of K7 in the Taurus star-forming  
region \citep{herbig88}. We presented its {\it Spitzer} IRS spectrum in \citet{furlan06} 
as part of the sample of Class III objects in Taurus, noting that it showed an infrared excess 
beyond about 12 $\mu$m (see Figure \ref{V819Tau_IRS}). However, we could not confirm 
whether this excess was real due to the presence of a ``companion'' 10\arcsec\ to the south 
that is detected in 2MASS images \citep{skrutskie06}. 
Two other weak-lined T Tauri stars in Taurus, HBC 427 and V410 X-ray 3, with spectral
types of K5 and M6, respectively \citep{steffen01,strom94}, were also introduced in
\citet{furlan06} as objects with uncertain infrared excesses. While HBC 427 has a
``companion'' star about 15\arcsec\ to the southeast that is seen in 2MASS images,
V410 X-ray 3 appears to be single.  
In the meantime, we obtained IRS peak-up images of V819 Tau and HBC 427, and we
re-reduced the IRS spectra of all three objects. We do not confirm the mid-infrared
excess in HBC 427 and V410 X-ray 3, but, together with Multiband Imaging Photometer 
for {\it Spitzer} \citep[MIPS;][]{rieke04} 24 $\mu$m images from the {\it Spitzer} 
archive, we establish that the infrared excess is intrinsic to V819 Tau, and we apply simple 
models to derive the distribution of dust around this T Tauri star. 

\newpage

\section{Observations}

\subsection{V819 Tau}

\begin{figure}
\plotone{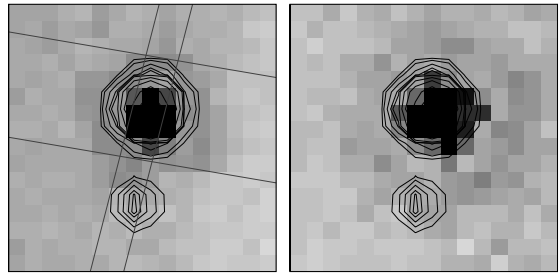}
\caption{IRS peak-up images ({\it left}: 16 $\mu$m; {\it right}: 22 $\mu$m) of V819
Tau, with superposed contours of the 2MASS $K_s$ image. North is up, east is to the left;
the plate scale amounts to 1$\farcs$8/pixel, and the field of view is 
29\arcsec$\times$29\arcsec. 
The SL and LL slit positions are also sketched in the 16 $\mu$m image; the SL slit is the 
narrower one running roughly north-south.
\label{V819Tau_peakup}}
\end{figure}

The IRS data of V819 Tau were taken on 2004 March 5 with the two low-resolution 
IRS modules (Short-Low [SL] and Long-Low [LL], 5.2--14 $\mu$m and 14--38 
$\mu$m, respectively, $\lambda$/$\Delta\lambda$ $\sim$ 90) in staring 
mode, and the spectrum was extracted and calibrated as described in \citet{furlan06}. 
The narrower SL slit (3.6\arcsec\ wide) was oriented roughly along the north-south 
direction, with the LL slit (10.5\arcsec\ wide) perpendicular to the SL slit (see a sketch 
of the slit positions in Figure \ref{V819Tau_peakup}). Since we extracted the object 
with a variable-width column that varies from a width of about 3 pixels to about 5 pixels 
from shortest to longest wavelengths in each module, the largest extraction radius in SL 
amounts to 4.5\arcsec, while in LL this value is $\sim$ 13\arcsec. These beams
are just slightly larger than the diffraction-limited sizes of a point-source at 14 and 
38 $\mu$m, respectively, observed with {\it Spitzer}. The companion star 10\arcsec\ 
to the south of V819 Tau is very faint at near-infrared wavelengths, with flux ratios 
relative to V819 Tau of 0.04-0.03 from $J$ to $K_s$; if it were a very red object, 
it could have contaminated some of the emission in LL, but not in SL.

To determine the mid-infrared flux of the ``companion'', we obtained IRS red (16 $\mu$m) 
and blue (22 $\mu$m) peak-up images on 2006 March 7 ({\it Spitzer} AORID 16269312). 
We dithered the target in two positions displaced by 9\arcsec\ from the center of the 
peak-up field of view; at each position we integrated for 6 seconds. For the red peak-up images, 
we performed this sequence twice. By subtracting two such nod pairs from each other, we 
removed any background emission; aperture photometry with a radius of 3 pixels and a sky 
annulus between 18 and 20 pixels yielded 16 and 22 $\mu$m fluxes of 18.4 $\pm$ 0.2  
and 24.6 $\pm$ 0.2 mJy, respectively, for V819 Tau. These flux values include aperture 
correction factors of 1.418 and 1.561 for the blue and red peak-up, respectively, and they
were derived assuming a spectral shape of ${\nu} F_{\nu}$=constant for the source
spectrum (see the document pu\_fluxcal.txt found on the IRS page of the Spitzer Science 
Center's website).

Figure \ref{V819Tau_peakup} shows the 16 and 22 $\mu$m peak-up images of V819 Tau; 
each represents the sum of two peak-up exposures and was rotated such that north is up 
and east is to the left. 
Superposed on Figure \ref{V819Tau_peakup} are the contours of the 2MASS $K_s$ image. 
While the southern ``companion'' is clearly present in 2MASS, it is not detected at 16 or 22  
$\mu$m. We derived upper limits for its flux from the rms in 3$\times$3 (for the blue 
peak-up) and 5$\times$5 (for the red peak-up) pixel boxes centered at its position in the 
co-added image pairs. We thus obtained 16 and 22 $\mu$m 5-$\sigma$ upper limits of 
0.8 and 2.0 mJy, respectively.

We also retrieved the MIPS 24 and 70 $\mu$m post-bcd, mosaicked images from the 
{\it Spitzer} archive (program 3584, PI D. Padgett; {\it Spitzer} AORID 11229696). 
By using a circular aperture with a radius of 14 pixels and a sky annulus between 16 and 
20 pixels, and by applying an aperture correction factor of 1.082, we measured a flux of 
22.3 $\pm$ 1.3 mJy at 24 $\mu$m for V819 Tau. At 70 $\mu$m, we used the 
filtered bcd image from the {\it Spitzer} archive and derived a 5-$\sigma$ upper limit 
of 30 mJy from the rms in a 13$\times$13 pixel box centered at the position of V819 
Tau. These fluxes assume a spectral shape of ${\nu} F_{\nu} \propto {\nu}^3$; 
corrections for different spectral shapes are minor, e.g., 4\% at 24 $\mu$m and 9\% at 
70 $\mu$m for ${\nu} F_{\nu}$=constant (see MIPS Data Handbook).

Thus, the imaging data confirms that the IRS spectrum of V819 Tau is not contaminated 
at any significant level by the source that is seen to its south at near-IR wavelengths. 
We will therefore consider the long-wavelength emission as coming only from the disk
around V819 Tau.

\subsection{HBC 427}

\begin{figure}
\plotone{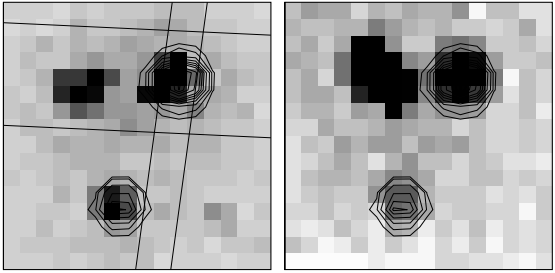}
\caption{As Figure \ref{V819Tau_peakup}, but for HBC 427. \label{HBC427_peakup}}
\end{figure}

Similar to V819 Tau, the IRS data of HBC 427 were taken on 2004 February 27 with 
the SL and LL modules in staring mode and reduced in the same manner.
The slits were oriented similarly to V819 Tau (see Figure \ref{HBC427_peakup}). 
In 2MASS, only two sources are detected; the southern ``companion'' is about half 
as bright as HBC 427 at $K_s$, but 15\arcsec\ from it ($\sim$ 8 \arcsec\ 
from the nearest edge of the LL slit), making it unlikely to contribute substantially 
to the LL flux. However, after we obtained IRS red and blue 
peak-up images of HBC 427 on 2006 March 8 ({\it Spitzer} AORID 16270336), we 
discovered an infrared-bright companion of HBC 427 that does not show up in 2MASS 
images. It lies $\sim$ 9\arcsec\ to the east of HBC 427 and is the brightest object 
of the three at 22 $\mu$m. This object, HBC 427/1, was actually already identified 
by \citet{massarotti05}, who imaged it at 1.025 $\mu$m and retrieved it from the 
USNO-B1.0 catalog. Our LL observation represents the combined flux of HBC 427 and 
HBC 427/1; in particular starting at about 20 $\mu$m, the latter source dominates 
the emission.

In order to measure the peak-up fluxes of HBC 427 and HBC 427/1, we performed
aperture photometry. Even though our observing sequence was the same as for
the peak-up observations of V819 Tau, we could not subtract the background by
subtracting the nod pairs, since the field was too crowded. Thus, the only sky
subtraction was performed during aperture photometry, where we used an aperture 
radius of 2 pixels and a sky annulus between 11-12 and 13-14 pixels. By applying
aperture correction factors of 1.661 and 1.885 for the blue and red peak-up, 
respectively, and assuming a spectral shape of ${\nu} F_{\nu}$=constant, 
we derived 16 $\mu$m fluxes of 11.0 $\pm$ 0.2 mJy and 7.2 $\pm$ 0.2 mJy 
for HBC 427 and HBC 427/1, respectively, and 22 $\mu$m fluxes of 7.0 $\pm$ 
0.2 mJy and 14.4 $\pm$ 0.3 mJy for HBC 427 and HBC 427/1, respectively.

We also measured the fluxes of these two objects in the MIPS 24 $\mu$m 
post-bcd, mosaicked image from the {\it Spitzer} archive (program 173, PI
N. Evans; {\it Spitzer} AORID 5683200). Applying a PSF fit using the star
HD 31305, which appears in the same image, as PSF calibrator, yielded
24 $\mu$m fluxes of 5.0 $\pm$ 1.3 mJy and 17.8 $\pm$ 1.2 mJy for 
HBC 427 and HBC 427/1, respectively. 

\begin{figure}
\centering
\includegraphics[angle=90, scale=0.35]{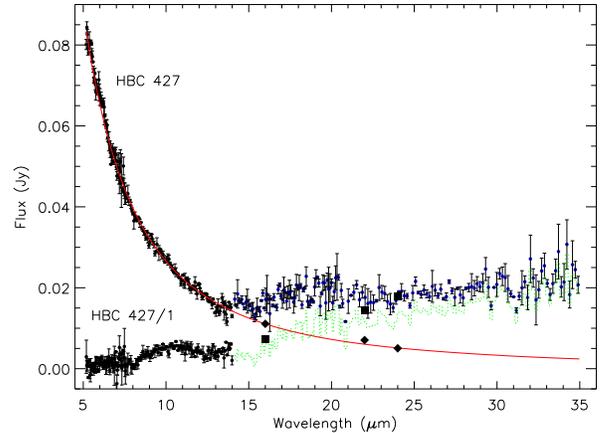}
\caption{IRS spectra of HBC 427 and HBC 427/1; the LL spectrum represent emission
from both objects. The filled diamonds and squares are the peak-up and MIPS 24 $\mu$m
fluxes of HBC 427 and HBC 427/1, respectively. The red solid line shows a blackbody fit to
the SL spectrum of HBC 427 (T=4310 K, $\Omega$=2.65 $\times$ 10$^{-19}$ sr). 
The green dotted line is an estimate of the LL spectrum of HBC 427/1, obtained by 
subtracting the blackbody fit from the combined LL spectrum. 
The data were dereddened using Mathis's reddening law for $R_V$=3.1 \citep{mathis90} 
and $A_V$=0.5 \citep{furlan06}. 
\label{HBC427_HBC427-1_IRS}}
\end{figure}

We obtained an IRS spectrum of HBC 427/1 on 2006 March 15 ({\it Spitzer} 
AORID 16262400) using SL and LL in staring mode. While we were able to extract and
calibrate the SL spectrum for HBC 427/1, the LL spectrum is again a combination
of the flux of HBC 427/1 and HBC 427. Since this LL spectrum has better signal-to-noise
than the one obtained earlier due to longer integration times, we use this spectrum
in the following plot. Figure \ref{HBC427_HBC427-1_IRS} shows the IRS spectra
of both HBC 427 and HBC 427/1, as well as the photometry we obtained from the
peak-up and MIPS images. A blackbody fit to the SL spectrum demonstrates that 
the 16, 22, and 24 $\mu$m fluxes of HBC 427 are essentially photospheric; the 
infrared excess is solely attributable to HBC 427/1.

\subsection{V410 X-ray 3}

\begin{figure}
\centering
\includegraphics[angle=90, scale=0.35]{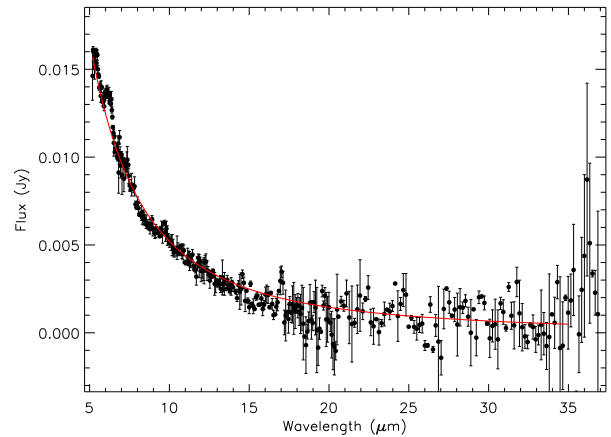}
\caption{IRS spectrum of V410 X-ray 3 and a blackbody fit to the SL spectrum
with T=3520 K and $\Omega$=6.62 $\times$ 10$^{-20}$ sr (red, solid line).
The data were dereddened using Mathis's reddening law for $R_V$=3.1 \citep{mathis90} 
and $A_V$=1.5 \citep{kraus06}. 
\label{V410Xray3_IRS}}
\end{figure}

The IRS data of V410 X-ray 3 were taken on 2004 Feb 7, also using SL and LL
in starting mode. Since this object is very faint in the first order of LL (20-36 $\mu$m), 
we first subtracted sky emission by using the off-order observation, then extracted its 
20-36 $\mu$m spectrum while also fitting a first order polynomial to the background 
emission in the slit. This procedure allowed for more accurate background subtraction. 
We compared reductions done with basic calibrated data products from the {\it Spitzer} 
Science Center's S11, S13.2, and S18.7 pipeline versions, and the differences 
were minimal. As a result of the improved background subtraction, we do not confirm 
the tentative excess seen in \citet{furlan06}; the IRS spectrum seems consistent with
photospheric emission alone (see Figure \ref{V410Xray3_IRS}).

\section{Spectral Energy Distributions}

\citet{skrutskie90} stated that V819 Tau has a spectral energy distribution
(SED) that identifies it as a ``transitional'' disk, i.e., a disk with an optically thin inner region
and an optically thick outer disk. \citet{beckwith90} detected millimeter-continuum emission
from V819 Tau and inferred a disk mass of 0.032 {\Msun}, implying that its outer disk 
is fairly massive. However, the far-IR SED of \citet{skrutskie90} had to rely on {\it IRAS} 
data, which are not reliable for such a faint source: according to \citet{weaver92}, the
{\it IRAS} 25 and 60 $\mu$m fluxes of V819 Tau are both 100 mJy, while our data shows
that the flux at 30 $\mu$m amounts to 20 mJy. In addition, later measurements by 
\citet{osterloh95} and \citet{duvert00} only determined an upper limit to the 1.3 mm 
flux of 9 and 2.5 mJy, respectively, and \citet{andrews05} could also only give upper 
limits to the 450 and 850 $\mu$m fluxes (317 and 9 mJy, respectively); the latter authors 
determined a disk mass of less than 0.0004 \Msun. Thus, it seems that the outer disk is 
not very massive. 

Since the more recent measurements are much more precise, they should be favored
when interpreting V819 Tau. Given that the excess emission from this T Tauri star is 
weak, earlier data with less sensitive instruments are more uncertain. 
The object 10\arcsec\ to the south of V819 Tau is even fainter; its low infrared flux 
values support the suggestion of \citet{woitas01} that it is likely a background star 
due to its low luminosity and its inferred age that places it well below the main sequence 
when plotted with evolutionary tracks.

Opposite to what we found for V819 Tau, the new {\it Spitzer} data of HBC 427 
show that the infrared excess seen in the long-wavelength part of the IRS spectrum 
is attributable to a nearby companion, HBC 427/1. 
Even though \citet{massarotti05} suggest it could be a hot background 
star based on its optical to near-IR colors, the infrared excess indicates that it is
an object surrounded by dust. We estimated the emission of HBC 427/1 beyond
14 $\mu$m by subtracting a blackbody fit to the SL spectrum of HBC 427 from
the LL spectrum (see Figure \ref{HBC427_HBC427-1_IRS}); the 16, 22, and
24 $\mu$m photometry of HBC 427/1 agrees reasonably well with this estimated 
spectrum. The infrared excess consists of continuum emission, as well as optically
thin radiation from silicate dust grains, whose signatures are broad 10 and 20 
$\mu$m emission features, similar to what is seen in protoplanetary disks
\citep[e.g.,][]{furlan06}.
However, without any additional data, we cannot conclusively determine the 
nature of this object.

The very low-mass star V410 X-ray 3 is fainter than V819 Tau or HBC 427; only a 
careful new reduction of its IRS spectrum showed that it does not have an infrared 
excess. A blackbody fit to the 5-14 $\mu$m spectrum also matches the 15-35 
$\mu$m fluxes (see Figure \ref{V410Xray3_IRS}). However, the spectrum 
beyond 20 $\mu$m is quite noisy; we can therefore not exclude excesses smaller 
than about 1 mJy in this wavelength range. \citet{luhman09} find that a very 
small excess of $\sim$ 0.4 mag may be present in MIPS 24 $\mu$m data, 
which is smaller than our upper limit of $\sim$ 1 mag for an excess at this 
wavelength. More sensitive observations at longer wavelengths are necessary 
to confirm this small excess; given the current data, we consider V410 X-ray 3 
as diskless.

\section{The IRS Spectrum and Model Fits of V819 Tau}

Our IRS spectrum confirms the transitional disk nature of V819 Tau, but it reveals that the
mid-infrared excess is much smaller than previously thought (see Figure \ref{V819Tau_IRS});
it sets in at about 8 $\mu$m, and $F_{\nu}$ is roughly constant with wavelength between 
23 and 33 $\mu$m. The very small infrared excess ($L_{IR}/L_{bol} \sim $ 10$^{-3}$) 
makes V819 Tau the most advanced transitional disk in the Taurus sample observed to
date. Also, as opposed to the other transitional disks, whose inner holes are delimited by a 
disk wall that gives rise to a steep continuum slope beyond 13 $\mu$m, the spectral index 
between 13 and 25 $\mu$m for V819 Tau is more negative, in a range typical for settled 
disks \citep{furlan06}.

V819 Tau is a K7 star \citep{walter81}, which suggests $T_{eff}$=4060 K 
\citep{kenyon95}; combined with a luminosity of 0.81 {\Lsun} (KH95), we derive a 
stellar radius of 1.8 {\Rsun}. Using evolutionary tracks computed by \citet{baraffe98}, 
we infer a stellar mass of 0.8 {\Msun} (as well as an age of 2 Myr), which indicates 
that the surface gravity is of the order $log(g) \sim$ 3.5-4.0. Thus, for the photosphere 
we adopt one of the NextGen photospheric models \citep{allard00} with 
$T_{eff}$=4000 K, $log(g)$=3.5, scaled by assuming a stellar radius of 
2 {\Rsun} and a distance of 140 pc \citep{bertout99}. 

\begin{figure}
\centering
\includegraphics[angle=90,scale=0.35]{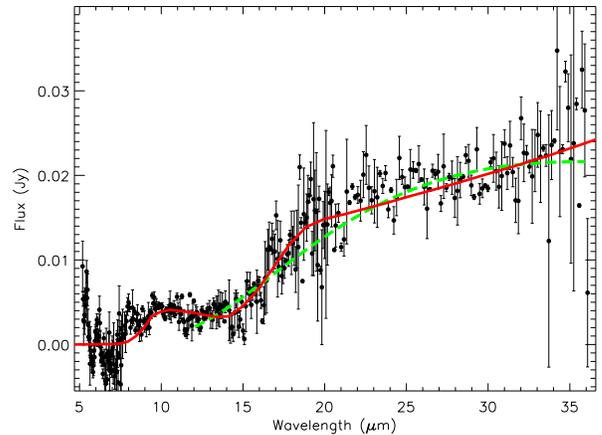}
\caption{Photosphere-subtracted IRS spectrum (see text for details) and fits
to the infrared excess: a blackbody fit from 12 to 36 $\mu$m ({\it dashed green
line}), yielding a temperature of 143 K and a solid angle of the emitting area of 
3.90 $\times$ 10$^{-16}$ sr, and an optically thin model fit ({\it solid red
line}). The latter model is one of several very similar fits that use amorphous 
olivines with grains ranging in size between 0.1 and 1 $\mu$m (up to 15 discrete 
sizes), distributed between 1 and several hundred AU; the surface density is 
constant with respect to radius and amounts to $\sim$ 4 $\times$ 
10$^{-7}$ g cm$^{-2}$ for all grains. 
\label{V819Tau_phot_sub}}
\end{figure}

The photosphere-subtracted IRS spectrum is shown in Figure \ref{V819Tau_phot_sub}. 
A single-temperature blackbody fit from 12 to 36 $\mu$m yields a blackbody 
temperature of 143 K. If the excess emission were optically thick, generated by a 
solid blackbody surface, this temperature would imply an inner disk radius of 6.8 AU. 
However, the low excess flux at longer wavelengths is not typical for optically thick 
emission. Assuming optically thin emission from gray grains in radiative equilibrium 
whose absorption cross section equals their geometric cross section (i.e, large, 
blackbody grains), the inner disk radius would lie at 3.4 AU. More realistically, 
the absorption efficiency varies with wavelength, as is the case for smaller grains
\citep[e.g.,][]{dorschner95}. We actually detect the signatures of small, amorphous 
grains of silicate composition (``amorphous silicates'') in an optically thin medium as 
weak, broad features centered at 10 and 20 $\mu$m \citep[e.g.,][]{jaeger94} 
in the photosphere-subtracted spectrum. If the dust grains are small (sub-$\mu$m), 
their absorption efficiency in the infrared is less than at shorter wavelengths, leading 
to so-called superheating, which would result in a grain being at a somewhat larger 
distance from the star than a blackbody at the same temperature.

The lack of substantial excess emission below about 8 $\mu$m suggests that the inner
disk regions are cleared of small dust grains. In order to derive the disk structure of V819 Tau, 
we applied an optically thin model fit to the infrared excess emission (8-34 $\mu$m). 
We adopted the optical constants of ``astronomical'' silicates from \citet{draine84} 
and chose different grain sizes with various surface density distributions (which act as a 
scaling factor for the emission) to match the observed excess. The grain size distribution 
was assumed to vary as a power law with an index of -3.5. At each distance from the 
star, we calculated the equilibrium temperature for the average grain size, and then 
integrated the optically thin emission from all grains over a certain radius range to 
achieve a good fit. 

A range of model parameters results in fits that reproduce the weak 10 and 20 
$\mu$m silicate emission features and rising continuum seen in the 
photosphere-subtracted spectrum: minimum grain sizes from 0.1 to 0.5 $\mu$m, 
maximum grain sizes of 0.5 or 1 $\mu$m, an inner disk radius of 1 AU,
and outer disk radii from 300 to 800 AU. Several models lie within reduced 
$\chi^2$ ($\chi^2_{\nu}$) values of 0.3 from the best-fit model 
($\chi^2_{\nu}$=19.1), which is shown in Figure \ref{V819Tau_phot_sub}. 
These models are basically indistinguishable except for slight differences beyond 
30 $\mu$m, where the flux is most sensitive to the outer disk radius; at 70
$\mu$m, they all lie at or below the upper limit derived from MIPS data.
A surface density constant in radius is necessary to reproduce the slowly rising 
20-35 $\mu$m excess, as is an outer radius of several 100 AU; if the disk 
were smaller than $\sim$ 100 AU, the model would still fit the excess out to 
20 $\mu$m, but increasingly underpredict the flux at longer wavelengths. 
An inner radius of 0.5 AU instead of 1 AU results in an increase in 
$\chi^2_{\nu}$ by about 1. Summarizing, we require sub-$\mu$m grains 
distributed uniformly between $\sim$ 1 AU and at least 100 AU to reproduce
the infrared excess of V819 Tau.
The mass of optically thin dust from our models amounts to up to 2.4 $\times$ 
10$^{26}$ g, or 3.2 lunar masses, for the largest disk sizes, and about 4.6 
$\times$ 10$^{25}$ g, or 0.6 lunar masses, for a disk of $\sim$ 400 AU 
radius. The optical depth of the emission by all grains is only 
$\sim$ 10$^{-4}$-10$^{-3}$ over the infrared spectral range.
  
\begin{figure}
\centering
\includegraphics[angle=90,scale=0.35]{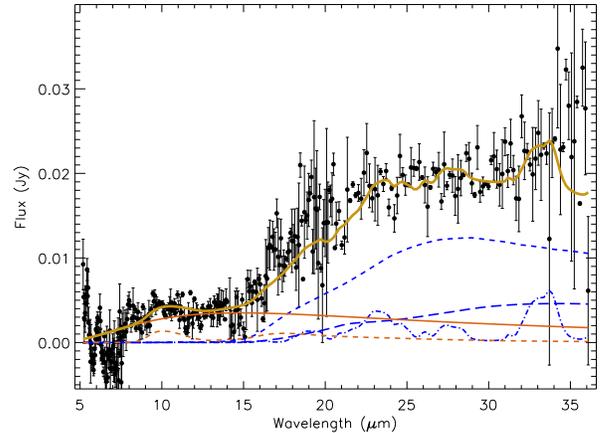}
\caption{IRS spectrum as in Fig.\ \ref{V819Tau_phot_sub}, but with an optically 
thin model fit aimed at deriving the dust composition. The orange lines represent the 
warm dust (345 K), the blue lines the cold dust (85 K); the dust components required 
for the fit are small amorphous olivine ({\it dashed lines}), large amorphous pyroxene
({\it long-dashed lines}), forsterite ({\it dash-dotted line}), and a blackbody
component ({\it solid, thin line}). The thick line is the sum of all components. 
\label{V819Tau_dust_comp}}
\end{figure}

\begin{deluxetable}{lll}
\tablewidth{\linewidth}
\tablecaption{Dust Components of V819 Tau \label{tab_dust}}
\tablehead{
\colhead{Dust component} & \colhead{T [K]} & \colhead{Mass [lunar masses]} 
}
\startdata
small amorphous olivine$^a$ & 345 & (2.0 $\pm$ 1.4) $\times$ 10$^{-6}$ \\
small amorphous olivine$^a$ & 85 & (9.9 $\pm$ 0.8) $\times$ 10$^{-3}$ \\
large amorphous pyroxene$^b$ & 85 & (3.1 $\pm$ 0.8) $\times$ 10$^{-3}$ \\
forsterite$^c$ & 85 & (1.2 $\pm$ 0.5) $\times$ 10$^{-3}$ \\
\tableline
blackbody$^d$ & 345 K & (4.5 $\pm$ 0.7) $\times$ 10$^{-18}$ sr 
\enddata
\tablecomments{
(a) Optical constants from \citet{dorschner95};
(b) optical constants from \citet{jaeger94}; 
(c) optical constants from \citet{sogawa06};
(d) the quantity listed in the third column is the solid angle.
}
\end{deluxetable}

In order to gauge the dust composition of the V819 Tau disk, we applied a model
to the photosphere-subtracted spectrum beyond 7.7 $\mu$m following the
methods of \citet{sargent06,sargent09a,sargent09b}. The best $\chi^2_{\nu}$
value of 14.7 is achieved by using a warm dust component at 345 K and a cold 
one at 85 K. The warm component includes a blackbody, which represents continuum 
emission from an optically thick medium or featureless emission from grains in an 
optically thin medium, and small amorphous silicates of olivine composition 
(``amorphous olivine''). The cold component is dominated by small amorphous 
olivine, with contributions of large amorphous silicates of pyroxene composition 
(``amorphous pyroxene'') and forsterite (see Table \ref{tab_dust} and Figure 
\ref{V819Tau_dust_comp}).
These model components represent a best fit to the data and were narrowed
down from a larger sample of dust components that included small and large
amorphous silicates, enstatite, forsterite, and silica, which are typically found
in protoplanetary disks (see \citet{sargent09b} for details). 
The uncertainties listed in Table \ref{tab_dust} actually amount to a few standard 
deviations \citep[see][]{sargent09b}; combined with the fact that a blackbody
cannot fit the data well (see Figure \ref{V819Tau_phot_sub}), the identification of 
amorphous silicates in our data is quite robust. However, the detection of forsterite is 
rather tentative, since the spectrum is more noisy beyond 20 $\mu$m.

The dust mass is dominated by cold, sub-$\mu$m amorphous olivine grains, which are
typical for the interstellar medium. The presence of some large (5 $\mu$m radius,
60\% vacuum porous) amorphous pyroxene grains, and possibly also forsterite, suggests
that some dust processing has taken place. Assuming a distribution of dust between
1 AU and several tens of AU, the optical depth amounts to $\lesssim$ 10$^{-3}$.
Thus, the dust we observe in V819 Tau is dominated by silicates in an optically thin 
medium.

\section{Discussion and Conclusions}

Of the three Class III objects in Taurus that showed tentative mid-infrared excess 
emission in \citet{furlan06}, we could confirm an infrared excess only in V819 Tau.
While V819 Tau is a single star, HBC 427 and V410 X-ray 3 are close binaries:
HBC 427 is a spectroscopic binary with a semi-major axis of about 0.03\arcsec, 
comprised of a K5 and an M2 star \citep{steffen01}; V410 X-ray 3 consists 
of an M6 and an M7.7 star separated by $\sim$ 0.05\arcsec\ \citep{kraus06}.
Neither object shows signs of accretion \citep{strom94, kenyon98, mohanty05}, 
and therefore both dust and gas have already dissipated in their inner disks. HBC 427
is also not detected at sub-millimeter wavelengths, implying an upper limit for the
disk mass of 0.0007 \Msun \citep{andrews05}. It is likely that interaction between
the binary and the disk in these systems is the cause for the absence of disk material
\citep{jensen94}.

V819 Tau is also not accreting any more; its H${\alpha}$ 10\% width amounts to 
166 km s$^{-1}$ \citep{nguyen09}, which is considerably less than the lower limit of 
270 km s$^{-1}$ that characterizes accretors \citep{white03}. Also the low 
H${\alpha}$ equivalent width of 1.7-3.2 {\AA} \citep{strom94, kenyon98}
confirms its nature as a weak-lined T Tauri star. \citet{white01} determined an 
upper limit of $1.4 \times 10^{-9}$ {\Msun} yr$^{-1}$ for the mass accretion 
rate based on the lack of $U$-band excess. No H$_2$ emission from warm inner disk 
regions was detected \citep{bary03}, implying that both the gas and dust have been
removed there. Our data indicate that there are no small, warm dust grains within 
approximately 1 AU from the star.

One process that can generate cleared inner disk regions is photoevaporation \citep{clarke01};
it sets in once the mass accretion rate has dropped to levels below a few 10$^{-10}$ 
{\Msun} yr$^{-1}$, and it acts by eroding the disk in a photoevaporative flow beyond 
a certain radius, which, for V819 Tau, amounts to 7.1 AU. This value is larger than 
the inner radius of our best-fit optically thin disk models, but it is remarkably close to the 
value of 6.8 AU we estimated by assuming the excess emission to arise from an optically
thick, blackbody surface. 
However, it seems that the dust emission is optically thin, given the presence of
10 and 20 $\mu$m silicate emission features and the low infrared excess luminosity 
($L_{IR}/L_{bol} \sim $ 10$^{-3}$). Moreover, our models require small 
dust grains at distances of at least $\sim$ 1 AU from the star. Thus, photoevaporation 
cannot explain the inner disk clearing of V819 Tau; a change in grain opacity caused 
by grain growth to sizes well beyond 1 $\mu$m in this region, or the gravitational 
influence of another body, such as a planet, are possible explanations for the inner
disk clearing.

The question also arises whether the material we detect around V819 Tau is primordial
or already second-generation dust. Detection of gas emission from the inner disk would 
support the primordial nature of the disk; so far, the absence of accretion signatures and 
H$_2$ emission from the inner disk suggest that the gas has already dissipated.
Without the gas, dust particles would be subject to collisions, Poynting-Robertson (PR)
drag, radiation pressure, and corpuscular stellar wind pressure \citep[e.g.,][]{wyatt99, 
chen06}, which, in the region from $\sim$ 1 to 100 AU from V819 Tau, would limit
the lifetime of sub-$\mu$m grains to $\lesssim$ 10$^5$ years. 

Therefore, in the absence of gas, dust would have to be continuously replenished, 
since the age of V819 Tau is about 2 Myr. V819 Tau would be one of the youngest 
debris disk systems; its infrared excess would lie on the higher side of what is usually
measured in debris disks \citep[e.g.,][]{chen06}, but still within the observed range. 
Collisions among planetesimals, as are thought to occur in debris disks, typically produce 
larger grains ($\gtrsim$ 10 $\mu$m), since the mid-infrared spectra of the majority 
of debris disks are featureless \citep{chen06,carpenter09}. The fact that small, 
amorphous silicate grains dominate the optically thin emission of V819 Tau, with just 
a minor possible presence of crystalline silicates, suggests that the dust we observe 
consists of mostly pristine material. Colliding or evaporating comet nuclei, which likely
form in the outer disk regions and therefore incorporate more unprocessed silicates 
into ice, could be the source of this pristine dust. 

It is also possible that we are just observing ``true'' primordial, unprocessed dust 
from large disk radii; this would imply that we are witnessing a rare stage in the life 
of a protoplanetary disk, when the gas has already been mostly removed from 
the inner disk, and the remaining dust is dissipated on a timescale of 10$^5$ 
years or less. Then, not only could the disk structure of V819 Tau be described 
as transitional due to the lack of infrared excess below $\sim$ 8 $\mu$m and 
the clear presence of such an excess beyond 12 $\mu$m, but also the evolutionary 
stage of this object could be considered transitional, i.e., having just transitioned from 
a primordial, optically thick disk to an optically thin one. 
Even though a few percent of the T Tauri population in Taurus can be considered
as transitional (defined as disks with substantial inner disk clearings and thus low
near-infrared excesses; \citealt{furlan09}), V819 Tau would stand out by its
very low infrared excess. It would be the only object in Taurus observed so far 
at an advanced transitional stage, when the disk material is already optically thin.
Given the small amount of dust left in the disk, planet formation would have to be 
already completed in this disk; this conclusion would also apply if the dust were 
second-generation and therefore the result of dynamically perturbed, remnant 
planetary building blocks.

\acknowledgments
We thank the referee for helpful comments that led us to expand and improve
this paper.
This work is based on observations made with the {\it Spitzer Space Telescope}, 
which is operated by the Jet Propulsion Laboratory (JPL), California Institute of 
Technology (Caltech), under NASA contract 1407. Support for this work was provided 
by NASA through contract number 1257184 issued by JPL/Caltech. E.F. was partly 
supported by a NASA Postdoctoral Program Fellowship, administered by Oak Ridge 
Associated Universities through a contract with NASA, and partly supported by NASA 
through the Spitzer Space Telescope Fellowship Program, through a contract issued 
by JPL/Caltech under a contract with NASA. 
This publication makes use of data products from the Two Micron All Sky Survey, 
which is a joint project of the University of Massachusetts and the Infrared Processing 
and Analysis Center/California Institute of Technology, funded by NASA and the NSF.
It has also made use of the SIMBAD and VizieR databases, operated at CDS (Strasbourg, 
France), NASA's Astrophysics Data System Abstract Service, and of the NASA/ IPAC 
Infrared Science Archive operated by JPL, Caltech, under contract with NASA.

Facilities: \facility{Spitzer(IRS)}

\end{document}